\RequirePackage[final]{graphicx}
\documentclass[aps, prr, superscriptaddress, twocolumn]{revtex4-2}
\usepackage[utf8]{inputenc}
\usepackage{amsmath}
\usepackage{amssymb}
\usepackage{amsfonts}
\usepackage{mathtools}
\usepackage{fixme}
\usepackage{color}
\usepackage{braket}
\usepackage{dsfont}
\usepackage{hyperref}
\hypersetup{
    colorlinks,
    citecolor=black,
    filecolor=black,
    linkcolor=blue,
    urlcolor=blue
}
\usepackage{url}
\usepackage{wasysym}

\usepackage{tikz}
\usetikzlibrary{calc}

\usetikzlibrary{positioning,arrows}
\usetikzlibrary{decorations}
\usetikzlibrary{decorations.pathmorphing,positioning}
\usetikzlibrary{decorations.markings}

\usepackage{tgtermes} 

\tikzset{square left brace/.style={ncbar=0.1cm}}
\tikzset{square right brace/.style={ncbar=-0.1cm}}

\definecolor{myred}{RGB}{214,26,70}
\definecolor{myreddark}{RGB}{76,8,38}
\definecolor{myblue}{RGB}{35,106,185}
\definecolor{mybluedark}{RGB}{19,56,99}
\definecolor{mybluebright}{RGB}{225,236,249}

\def\te{{\rm e}}

\def\bi{{\bf i}}
\def\bj{{\bf j}}

\def\bC{{\bf C}}

\def\bsigma{{\pmb \sigma}}

\def\nn{\nonumber}
\def\Re{{ \rm Re }}

\def\FM{{ \rm FM }}
\def\tr{{ \rm tr }}
\def\Ham{{ \hat{H} }}

\begin{document}
\title{Thermally induced localization of dopants in a magnetic spin ladder}
\date{\today}

\author{K. \ Knakkergaard \ Nielsen}
\affiliation{Max-Planck Institute for Quantum Optics, Hans-Kopfermann-Str. 1, D-85748 Garching, Germany}

\begin{abstract}
I unveil a novel variant of Anderson localization. This emergent phenomenon pertains to the motion of a dopant in a thermal spin lattice, rendered localized by thermal fluctuations. This is in stark contrast to the intrinsic origin of localization for quenched disorder. The system of interest consists of spin-$1/2$ particles organized in a two-leg ladder with nearest neighbor Ising interactions $J$. The motion of a hole -- the dopant -- is initialized by suddenly removing a spin from the thermal spin ensemble, which then moves along the ladder via nearest neighbor hopping $t$. I find that the hole remains \emph{localized} for all values of $J/t$ and for \emph{all} nonzero temperatures. The origin is an effective disorder potential seen by the hole and induced by thermal spin fluctuations. Its length scale is found to match with the underlying spin-spin correlation length at low temperatures. For ferromagnetic couplings ($J<0$), the associated localization length of the hole increases with decreasing temperature and becomes proportional to the correlation length at low temperatures, asymptotically delocalizing at low temperatures. For antiferromagnetic couplings ($J>0$), there is a smooth crossover between thermal localization at high temperatures to localization driven by the antiferromagnetic order at low temperatures. At infinite temperatures, the dynamics becomes independent of the sign of the spin coupling, whereby the localization length is a universal function of $|J|/t$, diverging as $(t/J)^{2}$ for $|J| \ll t$. Finally, I analyze a setup with Rydberg-dressed atoms, which naturally realizes finite range Ising interactions, accessible in current experimental setups. I show that the discovered localization phenomenon can be probed on experimentally accessible length- and timescales, providing a strong testing ground for my predictions. 
\end{abstract}

\maketitle

\section{Introduction} \label{sec.introduction}
The motion of dopants in magnetic spin lattices is crucial to our understanding of strongly correlated materials. The possible formation of polaronic quasiparticles and their induced interactions are believed \cite{Emery1987,Schrieffer1988,Dagotto1994} to be deeply connected to high-temperature superconductivity \cite{highTc}. Indeed, the behavior of such magnetic polarons in antiferromagnetic lattices \cite{Kane1989} has been shown to compare very well with exact diagonalization studies at zero temperature \cite{Martinez1991,Liu1991,Diamantis2021}. Furthermore, exciting new experiments has enabled the direct observation of dopant motion \cite{Ji2021}, made possible by the quantum simulation of Fermi-Hubbard-type models \cite{2010Esslinger,Boll2016,Zeiher2016,Cheuk2016b,Mazurenko2017,Hilker2017,Brown2017,Chiu2018,Brown2019,Koepsell2019,Chiu2019,Brown2020a,Vijayan2020,Hartke2020,Brown2020b,Koepsell2021,Gall2021} combined with single-site resolution techniques \cite{Bakr2009,Sherson2010,Haller2015,Yang2021}. The observed dynamics was successfully explained \cite{Nielsen2022_2} by the correlated formation and propagation of magnetic polarons \cite{Bohrdt2020}, in which the dopant eventually slows down and moves with a greatly reduced propagation speed. Despite these recent successes, there is still debate about the accuracy of this quasiparticle description \cite{Sheng1996,Wu2008,Zhu2015,White2015,Sun2019,Zhao2022}. 

The observation of such propagation dynamics \cite{Ji2021}, as well as the measurement of the spatial structures appearing around dopants \cite{Koepsell2019}, is a major new vantage point for our microscopic understanding of these systems. Indeed, previous work has mainly focused either on macroscopic observables such as currents driven by extrinsic force fields, spectroscopic measurements \cite{Kohstall2012,Hu2016,Jorgensen2016,Yan2020}, or Ramsey interferometry \cite{Cetina2015,Cetina2016,Schmidt2018,Skou2021,Skou2022}. While the measurement of currents gives invaluable insights into e.g. the physics of topological systems \cite{Thouless1983,Klitzing1986}, and spectral analyses gives access to some aspects of the appearing quasiparticles, it does not offer us detailed knowledge about their propagation. In particular, it does not provide a deep and microscopic understanding of the impact of the order -- or lack thereof -- of the environment. Recently, theoretical studies of dopant motion in thermal spin lattices \cite{Carlstrom2016,Nagy2017,Hahn2022} has ventured into this new paradigm. While some evidence of delocalization above the N{\'e}el temperature in a spin Ising environment \cite{Hahn2022} and hints of diffusive behavior at intermediate timescales at infinite temperatures has been seen for a more generic Fermi-Hubbard setup \cite{Carlstrom2016}, these studies were limited to fairly short evolution times and/or system sizes. As a result, the nature of the propagation on long timescales remains unsettled. 

\begin{figure}[t!]
\begin{center}
\includegraphics[width=1.0\columnwidth]{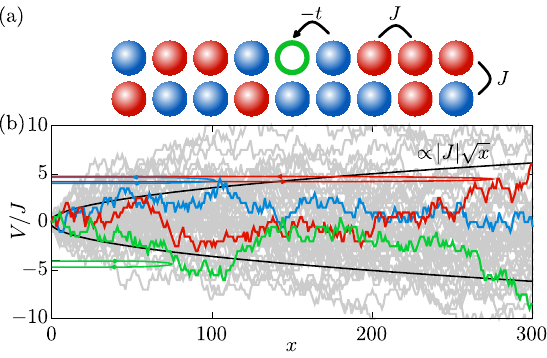}
\end{center}\vspace{-0.5cm}
\caption{(a) Motion of a single hole (green circle) along a spin ladder consisting of spin-$\uparrow$ (red balls) and spin-$\downarrow$ (blue balls), with hopping $t$. The spins are assumed to feature Ising-type nearest neighbor spin couplings $J$. (b) The hole experiences an emergent disorder potential $V$ due to the thermal spin fluctuations. In (b), $50$ realizations of this potential at infinite temperatures is shown as a function of the distance $x$ of the hole to its origin (grey lines), of which three are highlighted in color (red, blue, green). The variance of the potential grows as $|J|\sqrt{x}$ (black lines). As a result, no matter the realization, the hole will eventually backscatter and become \emph{Anderson} localized (colored lines with arrows). Here, this is shown for a hole with initial kinetic energy of $t = 4J$.}
\label{fig.introfig} 
\vspace{-0.25cm}
\end{figure} 

Intrigued by these investigations, I study the motion of a dopant in a thermal Ising spin ensemble. In particular, I consider a mixed-dimensional model in a two-leg ladder \cite{Bohrdt2022,Hirthe2023,Nielsen2023_1}. Here, the doped hole is allowed to move \emph{only along} the ladder with nearest neighbor hopping $t$, while the spin-$1/2$ particles are assumed to couple with Ising-type nearest neighbor spin couplings $J$ [Fig. \ref{fig.introfig}(a)]. I investigate both ferro- ($J<0$) and antiferromagnetic ($J>0$) couplings. The non-equilibrium motion of a hole at zero temperature in these two scenarios features highly distinct behaviors. Indeed, while the hole for antiferromagnetic couplings is localized due to a confining string potential \cite{Nielsen2023_1}, it moves completely freely in the ferromagnetic phase. However, at any nonzero temperatures I find that the hole experiences an emergent disorder potential, which localizes the hole at any value of $J/t$ and at \emph{any nonzero} temperature. This realizes a novel, \emph{emergent} type of Anderson localization \cite{Anderson1958,Fouque1999,Lagendijk2009}, as it is the backscattering of the hole on the disorder potential [Fig. \ref{fig.introfig}(b)] that leads to its localization. I emphasize that the origin of the localization does not come from \emph{quenched} disorder from, e.g., a random distribution of onsite energies \cite{Anderson1958}, but is an \emph{emergent} property of the system itself \cite{DeRoeck2014} arising at nonzero temperatures due to thermally induced spin fluctuations. 

The present studies, hereby, establishes rare insights into how the disorder in the underlying spin lattice crucially impacts the motion of dopants. Moreover, it describes an abrupt change in their qualitative characteristics. Indeed for ferromagnetic spin couplings, the dopant behaves as a free particle at zero temperature, but as soon as the phase transition at $T = 0$ is crossed, it completely looses its quasiparticle character, and its motion becomes localized. In this connection, it is interesting to note that absence of ballistic motion has been seen in similar systems with impurities and/or spin excitations \cite{Zvonarev2007, Zvonarev2009, Kantian2014, Kamar2019}. While such behavior can arise due to a quasiparticle breakdown \cite{Kantian2014, Kamar2019} of the impurities, such systems still support slow, subdiffusive, transport. In this regard, the total absence of transport found in the current model in of itself also seems to be rare effect, which strongly hinges on the correlated motion of the dopant with the underlying physical spins.

Finally, to showcase the possibility of detecting this localization phenomenon experimentally, I analyze a setup with Rydberg-dressed atoms in a two-leg optical lattice that supports finite-range Ising-type interactions \cite{Henkel2010}, already demonstrated experimentally \cite{Zeiher2016}. Here, I find that the localization can be probed on realistic timescales and system sizes, providing a strong testing ground for the predicted results.

The currently discovered localization effect would strictly speaking occur simultaneously with regular Anderson localization in one dimension for any realistic medium that would feature a nonzero disorder strength. In this context, I stress that the localization length due to thermal spin fluctations found in the present analysis should very easily be orders of magnitude smaller than the one arising due to Anderson localization, and, therefore, completely dominate the phenomenology. The computation of these results rests on a combination of two precise approaches. First, for a specific spin realization, I determine numerically exactly the non-equilibrium hole motion. Second, using large-scale Monte Carlo sampling of the thermal ensemble, I determine the appropriate thermal average of these pure state evolutions.

The Article is organized as follows. In Sec. \ref{sec.setup}, I describe the overall setup, including a description of the system Hamiltonian, as well as the thermal initial state of the spin ensemble, and finally the exact computation of the non-equilibrium hole motion for a specific spin realization in Sec. \ref{subsec.probability_amplitudes}. In Sec. \ref{sec.infinite_temperature}, I describe the universal regime of infinite temperatures. In Sec. \ref{sec.finite_temperature}, I go away from this universal limit and give a detailed analysis of the propagation across a wide range of temperatures. In Sec. \ref{sec.analysis_of_localization}, I give qualitative arguments for the dependencies on spin coupling and temperature seen in Secs. \ref{sec.infinite_temperature} and \ref{sec.finite_temperature}. In Sec. \ref{sec.Rydberg_experiment}, I finally analyze the Rydberg-dressed atoms setup, before I conclude in Sec. \ref{sec.conclusions}. Throughout the Article, I work in units where the reduced Planck constant, $\hbar$, and the lattice spacing is set to $1$.

\section{Setup} \label{sec.setup}
I consider a system of spin-$1/2$ particles placed along a two-leg ladder, described by a $t$-$J$ model with nearest neighbor \emph{Ising} interactions, 
\begin{equation}
\Ham = -t\sum_{\braket{\bi,\bj}_\parallel,\sigma} \left[\tilde{c}^\dagger_{\bi\sigma}\tilde{c}_{\bj\sigma} + \tilde{c}^\dagger_{\bj\sigma}\tilde{c}_{\bi\sigma}\right] + J \sum_{\braket{\bi,\bj}} \hat{S}^{(z)}_\bi\hat{S}^{(z)}_\bj.
\label{eq.Hamiltonian}
\end{equation}
The hopping is constrained through the operator $\tilde{c}^\dagger_{\bi\sigma} = \hat{c}^\dagger_{\bi\sigma}(1 - \hat{n}_\bi)$, such that at most a single spin resides on each site. I, furthermore, assume a mixed-dimensional setup in which the spins are only allowed to hop along the ladder. I will both analyze antiferromagnetic ($J > 0$) and ferromagnetic ($J < 0$) spin-coupling cases. To have an efficient description of the hole and spin excitation degrees of freedom, I employ a Holstein-Primakoff transformation on top of the \emph{ferromagnetic} ground state $\ket{\FM} = \ket{..\uparrow\uparrow\uparrow..}$, with all spins pointing up. As a result, the Hamiltonian $\Ham = \Ham_t + \Ham_J$ may be written in terms of the hopping,
\begin{align}
\Ham_t = t \sum_{\braket{\bi,\bj}_\parallel} \! &\Big[ \hat{h}^\dagger_{\bj} F(\hat{h}_{\bi}, \hat{s}_{\bi}) F(\hat{h}_{\bj}, \hat{s}_{\bj}) \hat{h}_{\bi} \nn \\
&+ \hat{h}^\dagger_{\bj}\hat{s}^\dagger_\bi F(\hat{h}_{\bi}, \hat{s}_{\bi}) F(\hat{h}_{\bj}, \hat{s}_{\bj}) \hat{s}_\bj\hat{h}_{\bi} \Big] + {\rm H.c.},
\label{eq.H_t_holstein_primakoff}
\end{align}
and the spin coupling
\begin{align}
\hat{H}_J =  J\sum_{\braket{\bi,\bj}} \Big[\frac{1}{2} \!-\! \hat{s}^\dagger_{\bi}\hat{s}_{\bi}\Big]\Big[\frac{1}{2} \!-\! \hat{s}^\dagger_{\bj}\hat{s}_{\bj}\Big] \Big[1 \!-\! \hat{h}_{\bi}^\dagger \hat{h}_{\bi}\Big] \Big[1 \!-\! \hat{h}_{\bj}^\dagger \hat{h}_{\bj}\Big].
\label{eq.H_J_holstein_primakoff}
\end{align}
Here, the spin excitation operator $\hat{s}^\dagger_\bi$ is bosonic, and creates a spin-$\downarrow$ on site $\bi$. Also, the hole is created by the operator $\hat{h}^\dagger_\bi$, and inherits the statistics of the underlying spins, be it fermionic \emph{or} bosonic \cite{Nielsen2023_1}. In the hopping Hamiltonian $\Ham_t$, the operator $F(\hat{h}, \hat{s}) = \sqrt{1 - \hat{s}^\dagger\hat{s} - \hat{h}^\dagger\hat{h}}$ ensures the single-occupancy constraint. The two terms in the bracket of $\Ham_t$ describe distinct hopping events. The first term describes a hole hopping from site $\bi$ to $\bj$ in the absence of a spin excitation on site $\bj$. The second term, on the contrary, describes this hopping in the presence of a spin excitation, whereby the hole and spin excitation swap places. While the Holstein-Primakoff transformation slightly complicates the expression for the Hamiltonian, it makes it much easier to write down concise expressions for the non-equilibrium wave functions to come. 

I assume that the system is closed and initially thermalized in the Gibbs state of the spins $\hat{\rho}_J = e^{-\beta \Ham_J} / Z_J$, i.e. in the absence of holes. Note, however, that I make no assumptions about how thermal equilibrium is established. One particular scenario would be via a controllable coupling to an external heat bath. Once the system has reached thermal equilibrium in the steady state, the coupling to the bath could be shut off. The resulting partition function $Z_J = \tr[e^{-\beta \Ham_J}]$ along with the spin-spin correlator, 
\begin{align}
C_z(d) &= 4 \braket{\hat{S}^{(z)}_{\bi}\hat{S}^{(z)}_{\bi + d \hat{x}}}_0 \nn \\
&= C_z^{(1)} e^{-d/\xi_1(\beta J)} + C_z^{(2)} e^{-d/\xi_2(\beta J)},
\end{align}
is derived analytically in Appendix \ref{app.thermodynamics}, where explicit expressions for the coefficients $C_z^{(i)}$ are also given. I, here, use the transfer matrix formalism \cite{Kramers1941} originally used for the Ising chain \cite{Ising1925} to the present two-leg ladder. As required by the one-dimensional geometry, the system is disordered at any temperature $T = (k_B\beta)^{-1}$. While there are two correlation lengths, I find that $\xi_1(\beta J) > \xi_2(\beta J)$ for any temperature. This correlation length,
\begin{align}
\!\!\!\!\xi_1(\beta J) &= \Bigg[ -\frac{\beta|J|}{4} + \ln \Bigg( \coth\frac{1}{2}\beta |J| \, \cosh \frac{1}{4}\beta |J|  \nn \\
&+ \sqrt{\left(\coth \frac{1}{2}\beta |J| \, \cosh \frac{1}{4}\beta |J|\right)^2 - 1}\Bigg)\Bigg]^{-1}\!\!,\!\!
\label{eq.correlation_length}
\end{align}
therefore, sets an essential length scale in the system at finite temperatures. In Eq. \eqref{eq.correlation_length}, $\cosh(x)$ and $\coth(x)$ are the hyperbolic cosine and cotangent, respectively. The non-equilibrium quench dynamics is now initialized by suddenly removing the spin at the origin $\bi = {\bf 0}$, leading to the initial density matrix
\begin{equation}
\hat{\rho}(\tau = 0) = \sum_{\sigma_0} \hat{c}_{{\bf 0}\sigma_0} \hat{\rho}_J \hat{c}^\dagger_{{\bf 0}\sigma_0} = \hat{h}^\dagger_{\bf 0} \hat{\rho}_J \hat{h}_{\bf 0} + \hat{h}^\dagger_{\bf 0}\hat{s}_{\bf 0} \hat{\rho}_J \hat{s}^\dagger_{\bf 0}\hat{h}_{\bf 0},
\end{equation}
where $\sigma_0 = \uparrow,\downarrow$ designates the spin configurations at the origin, and the latter expression rephrases it in terms of hole and spin-excitation operators. This setup is analogous to the situation studied in Ref. \cite{Hahn2022} in the two-dimensional Ising antiferromagnet. I stress, however, that while they find some evidence that the hole deconfines from its initial position somewhat above the N{\'e}el temperature, I, on the contrary, find that the hole is localized even in such a disordered phase, and both for ferro- and antiferromagnetic spin couplings. A plausible reason for this discrepancy is that Ref. \cite{Hahn2022} considers quite strong hopping amplitudes $t \geq J$, and system sizes of about $10\times10$, which is likely too small to distinguish delocalization from a localized though highly spread out hole at these high hopping amplitudes. It is, however, also possible that a hole moving in two dimensions will undergo diffusive or subdiffusive propagation on long timescales, and further studies should be carried out to settle this question. 

Since the system is assumed to be closed, the ensuing dynamics is unitary, i.e. $\hat{\rho}(\tau) = e^{-i\hat{H}\tau} \hat{\rho}(\tau = 0) e^{+i\hat{H}\tau}$. Expressing the density operator in the Ising basis with spin configurations $\bsigma$, this, hereby, allows us to write the time-evolved density matrix as the Boltzmann-weighted sum of pure-state time evolutions
\begin{equation}
\hat{\rho}(\tau) = \sum_{\sigma_0,\bsigma} \frac{e^{-\beta E_J(\sigma_0,\bsigma)}}{Z_J} \ket{\Psi_{\bsigma}(\tau)}\bra{\Psi_{\bsigma}(\tau)},
\label{eq.non_equilibrium_density_matrix}
\end{equation}
where $E_J(\sigma_0,\bsigma)$ is the magnetic energy of the spin realization $\sigma_0,\bsigma$ before the hole is introduced. With the hole and spin excitation operators at hand, we may express the non-equilibrium pure states $\ket{\Psi_{\bsigma}(\tau)}$ quite concisely. For a spin realization $\bsigma$, subsets $S_{\bsigma}^1, S_{\bsigma}^2$ of the sites in the first and second leg will have spins pointing down. Writing $\bi = l,j$ in terms of the legs $l = 1,2$ and site number along the leg $j$, the initial wave function can be expressed as 
\begin{equation}
\ket{\Psi_{\bsigma}(\tau = 0)} = \hat{h}^\dagger_{1,0} \prod_{j\in S_{\bsigma}^1} \hat{s}^\dagger_{1,j} \prod_{j\in S_{\bsigma}^2} \hat{s}^\dagger_{2,j} \ket{\FM}.
\end{equation}
As the hole starts to move along the ladder, the spins in leg 1 can be moved by a single lattice, while the spins in leg 2 remain stationary. Therefore, the state at any later time $\tau$ is
\begin{align}
\ket{\Psi_{\bsigma}(\tau)} \!=\! \Bigg[
  &\sum_{x \geq 0} \! C_\bsigma(x,\tau) \hat{h}^\dagger_{1,x} \!\!\!\!\prod_{\substack{j \in S_{\bsigma}^1 \\ 0 \leq j\leq x}} \!\!\!\!\hat{s}^\dagger_{1,j-1} \!\!\!\prod_{\substack{j \in S_{\bsigma}^1 \\ j > x}} \!\!\!\hat{s}^\dagger_{1,j} \nn \\
+ &\sum_{x < 0} \! C_\bsigma(x,\tau) \hat{h}^\dagger_{1,x} \!\!\!\! \prod_{\substack{j \in S_{\bsigma}^1 \\ x \leq j < 0}} \!\!\!\! \hat{s}^\dagger_{1,j+1} \!\!\!\prod_{\substack{j \in S_{\bsigma}^1 \\ j < x}} \!\!\! \hat{s}^\dagger_{1,j}\Bigg]\!\!\prod_{j \in S_{\bsigma}^2} \!\!\!\hat{s}^\dagger_{2,j}\!\ket{\FM}.
\label{eq.Psi_sigma}
\end{align}
The upper (lower) line describes that the spin excitations are moved by one site to the left (right), if the hole has passed it, and otherwise it remains where it was. Crucially, the probability amplitude to find the hole at site $x$ and time $\tau$ for a given spin realization $\bsigma$ only depends on these three variables, since the spin background is static. Moreover, the probability to observe the hole at position $x$ after time $\tau$ is simply the thermal average of the probabilities $|C_\bsigma(x,\tau)|^2$, 
\begin{align}
P(x,\tau) &= \tr\left[\hat{h}^\dagger_{1,x} \hat{h}_{1,x} \hat{\rho}(\tau)\right] \nn \\
&= \sum_{\sigma_0,\bsigma} \frac{e^{-\beta E_J(\sigma_0,\bsigma)}}{Z_J} |C_\bsigma(x,\tau)|^2.
\label{eq.prob_hole_dynamics}
\end{align}
In this manner, the problem of describing the motion of the hole has now been reduced to finding the probability amplitudes $C_\bsigma(x,\tau)$ for a given spin realization $\bsigma$ and then performing the sum in Eq. \eqref{eq.prob_hole_dynamics}. While this is hardly feasible to do exactly, I employ a standard Metropolis-Hastings algorithm \cite{Metropolis1953,Hastings1970} to perform accurate sampling of the sum, from which the root-mean-square distance is calculated
\begin{equation}
x_{\rm rms}(\tau) = \left[\sum_{x} x^2 P(x,\tau) \right]^{1/2}.
\label{eq.rms_distance}
\end{equation}

\subsection{Determining the probability amplitudes} \label{subsec.probability_amplitudes}
\begin{figure}[t!]
\begin{center}
\includegraphics[width=0.8\columnwidth]{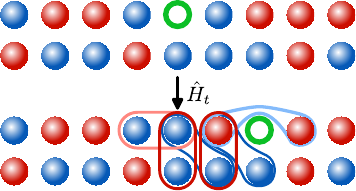}
\end{center}\vspace{-0.5cm}
\caption{As the hole moves through the ladder (top to bottom), it breaks up spin bonds across (dark blue) and along (light blue) the ladder. In the same manner new spin bonds are created across (dark red) and along (light red) the ladder. The effective magnetic potential experienced by the hole, hereby, arises by subtracting the energy of the broken spin bonds and adding the energy of the newly formed.}
\label{fig.explain_potential} 
\vspace{-0.25cm}
\end{figure}

In this subsection, I describe how the probability amplitudes $C_\bsigma(x,\tau)$ are determined numerically exactly, by which we can accurately describe the motion of the hole at essentially any temperature.  The key insight is that the structure of the states in Eq. \eqref{eq.Psi_sigma} lead to a very simplistic set of equations of motion,
\begin{align}
i\partial_\tau C_\bsigma(x,\tau) =&\,  V_\bsigma(x) C_\bsigma(x,\tau) \nn \\
& + t\left[C_\bsigma(x-1, \tau) + C_\bsigma(x+1,\tau)\right].
\label{eq.equations_of_motion}
\end{align}
Here, $V_\bsigma(x)$ designates the magnetic potential experiences by the hole as it moves through the lattice. This arises because motion of the hole changes the magnetic energy of the underlying spin lattice. Put in another way, as the hole moves through the ladder it breaks up a series of spin bonds and creates new ones as illustrated in Fig. \ref{fig.explain_potential}. The effective potential, $V_{\bsigma}(x) = V_{\bsigma,\parallel}(x) + V_{\bsigma,\perp}(x)$, can be decomposed in terms of 
an intra-leg potential
\begin{align}
V_{\bsigma,\parallel}(x) &= J[\sigma_{1,1}\sigma_{1,-1} - \sigma_{1,x}\sigma_{1,x + 1}], \; x > 0, \nn \\
V_{\bsigma,\parallel}(x) &= J[\sigma_{1,1}\sigma_{1,-1} - \sigma_{1,x}\sigma_{1,x - 1}], \; x < 0. 
\label{eq.intraleg_potential}
\end{align}
and a trans-leg potential
\begin{align}
V_{\bsigma,\perp}(x) &= J \sum_{j = +1}^x \sigma_{1,j}[\sigma_{2,j - 1} - \sigma_{2,j}], \; x > 0, \nn\\
V_{\bsigma,\perp}(x) &= J \sum_{j = -1}^x \sigma_{1,j}[\sigma_{2,j + 1} - \sigma_{2,j}], \; x < 0. 
\label{eq.transleg_potential}
\end{align}
Here, the index of the spins $\sigma = \pm 1/2 \equiv \,\uparrow,\downarrow$ refer to their positions \emph{before} the hole has started to move. In Eq. \eqref{eq.intraleg_potential}, the term $J \sigma_{1,1}\sigma_{1,-1}$ refers to the spin-bond energy arising around the origin as the hole has moved, while the term $J\sigma_{1,x}\sigma_{1,x + 1}$ for $x > 0$ is the energy of the bond broken up by the hole once it has moved to position $x$. 
These are shown in light red and light blue in Fig. \ref{fig.explain_potential}. Similarly, the two terms in the summand of Eq. \eqref{eq.transleg_potential} correspond to the energies $J\sigma_{1,j}\sigma_{2,j - 1}$, $J\sigma_{1,j}\sigma_{2,j}$ of the newly established and broken bonds every time the hole hops, shown in dark red and dark blue in Fig. \ref{fig.explain_potential}. 

The equations of motion in Eq. \eqref{eq.equations_of_motion} may actually be solved exactly by a Fourier transform [see Appendix \ref{sec.recursive_solution}]. However, the accompanying Fourier transform back to the time domain makes this computation less efficient than applying an exact diagonalization method. To set this up, we use Eq. \eqref{eq.equations_of_motion} to define the effective Hamiltonian $\mathcal{H}_\bsigma$ for a given spin realization $\bsigma$ with the matrix elements
\begin{align}
\mathcal{H}_\bsigma(x,x) &= V_\bsigma(x), \nn \\ 
\mathcal{H}_\bsigma(x\pm 1,x) &= \mathcal{H}_\bsigma(x,x\pm 1) = t.
\end{align}
By vectorizing the components $C_\bsigma(x,\tau)$ into $\bC_\bsigma(\tau)$, I obtain the time-evolution of the probability amplitudes by computing
\begin{equation}
\bC_\bsigma(\tau) = \te^{-i\mathcal{H}_\bsigma \tau}\bC_\bsigma(0),
\label{eq.vectorized_solution}
\end{equation}
with the initial condition that the hole starts out at $x = 0$: $\bC_\bsigma(x=0,\tau=0) = 1$. I compute this using the Python function "expm\_multiply" in the "scipy.sparse.linalg" package. By taking into account the sparseness of $\mathcal{H}_\bsigma$, and the fact that its size is only quadratic in the system size, this approach is highly efficient and allows system sizes of at least $20.000$ sites long. 

\section{Infinite temperature limit} \label{sec.infinite_temperature}
\begin{figure}[t!]
\begin{center}
\includegraphics[width=0.95\columnwidth]{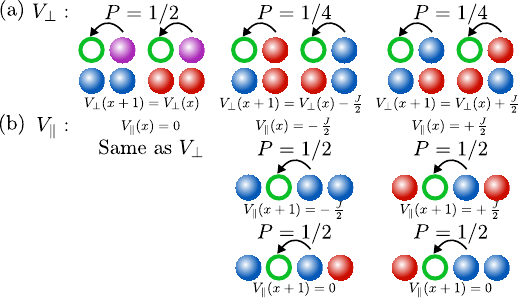}
\end{center}\vspace{-0.5cm}
\caption{At infinite temperature, the magnetic potential is random. (a) In each hop (arrows), the potential across the ladder remains unchanged with probability $P = 1/2$ (left), or goes down (middle) or up (right) by $J / 2$ with probability $P = 1/4$. A purple ball indicates that it does not matter, whether that spin is $\uparrow$ or $\downarrow$. (b) The potential along the ladder, $V_\parallel(x)$, can only take the values $0,\pm J / 2$, and the change depends on whether $V_\parallel(x) = 0$ (left), $V_\parallel(x) = -J / 2$ (middle), or $V_\parallel(x) = J / 2$ (right).}
\label{fig.random_ensemble} 
\vspace{-0.25cm}
\end{figure}

In the limit of infinite temperature, $\beta J = J / k_BT \to 0$, the partition function simply becomes the number of spin configurations $Z_J = 2^{2N}$, where $N$ is the number of sites in each leg. Furthermore, the terms in Eq. \eqref{eq.prob_hole_dynamics} all have the same statistical weight
\begin{align}
P(x,\tau) \to \frac{1}{2^{2N-1}}\sum_{\bsigma} |C_\bsigma(x,\tau)|^2.
\label{eq.prob_hole_dynamics_infinite_temp}
\end{align}
As a result, we need to describe how the hole moves in a completely \emph{random} spin ensemble. As was previously noticed in the context of Bethe lattices \cite{Bohrdt2020}, the resulting potential experienced by the hole, $V_{\bsigma}(x)$, becomes a disordered potential. In fact, in any hop, the potential changes \emph{at random} by an amount $|J| / 2$. This is detailed in Fig. \ref{fig.random_ensemble}. The equations of motion in Eq. \eqref{eq.equations_of_motion} now becomes very reminiscent of the 1D Anderson model for \emph{Anderson localization} \cite{Anderson1958}. However, in contrary to the original model, the potential is correlated from site to site, as is also apparent from Fig. \ref{fig.random_ensemble}, \emph{and} the trans-leg potential $V_{\bsigma,\perp}(x)$ becomes arbitrarily large at large $x$. This is in contrast to the usual case studied in Anderson localization, where some constant width of for the disordered potential is usually assumed. In fact, the potential performs a classical random walk in its allowed values. As a result, its variance, 
\begin{equation}
{\rm Var}[V_\bsigma(x)] = \frac{J^2}{8}\left[|x| + 1\right],
\label{eq.variance_potential}
\end{equation}
scales linearly in $|x|$, as shown explicitly in Appendix \ref{app.disordered_potential}. Due to the differences with the usual Anderson model, it is a priori not clear whether the hole will localize or not in this specific kind of disorder potential. By closer inspection of the probabilitic behavior of the potential sketched in Fig. \ref{fig.random_ensemble}, it becomes clear that the behavior at infinite temperature does not depend on the sign of the spin coupling, $J$, and the motion of the hole becomes universal in this limit.

\begin{figure}[t!]
\begin{center}
\includegraphics[width=1.0\columnwidth]{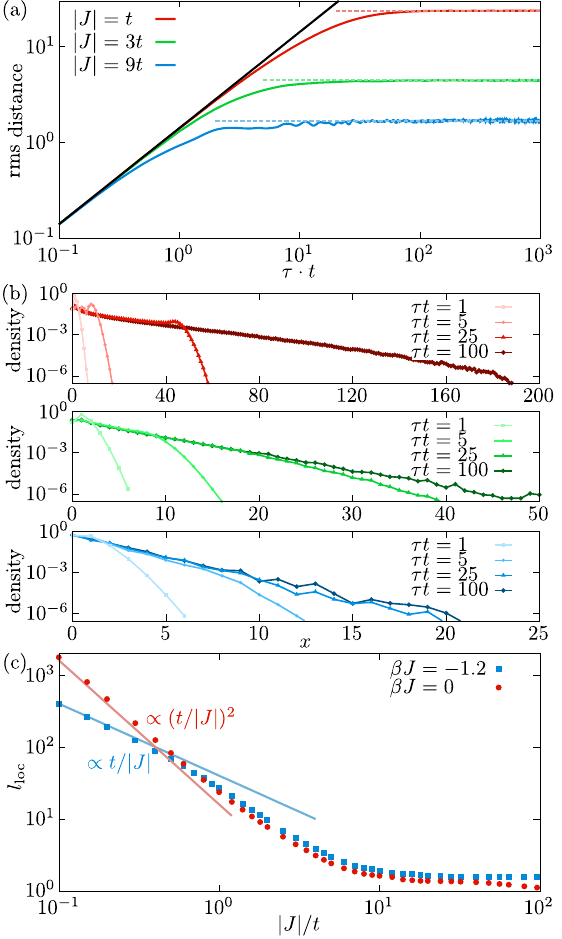}
\end{center}\vspace{-0.5cm}
\caption{(a) Rms distance of the hole versus time for indicated values of the spin coupling on a log-log scale. This shows an initial ballistic behavior with expansion speed $\sqrt{2}t$ (black line), before being localized on long timescales (dashed lines). (b) Hole density $P(x,\tau)$ at indicated times for the same values of $|J|/t$ as in (a), showing exponential localization of the hole. (c) Extracted localization length $l_{\rm loc}$ as the long-time asymptote of the rms distance as a function of $|J|/t$ at infinite temperature (red dots) compared to finite temperature (blue squares). For large spin couplings, $l_{\rm loc}$ approaches a nonzero value, whereas small spin couplings leads to distinct power-law behaviors at infinite [$\propto (t/J)^2$] and finite [$\propto t/|J|$] temperatures. The estimated statistical errors on the sampling are smaller than the linewidth/point size and are omitted.}
\label{fig.rms_distance_inf_temp} 
\vspace{-0.25cm}
\end{figure} 

The only remaining parameter in the system is $|J|/t$. For a given value of this ratio, I, thus, generate $N_\bsigma = 2000$ samples by using the probabilistic update rules for the potential shown in Fig. \ref{fig.random_ensemble}. For each of the generated realizations, I compute $C_\bsigma(x,\tau)$ up to very large times and from there the rms distance (Eq. \eqref{eq.rms_distance}). An example of the rms distance dynamics is given in Fig. \ref{fig.rms_distance_inf_temp}(a) for three indicated value of the spin coupling. For all of these, we clearly see that the hole remains \emph{localized}, stalling at a finite distance to its origin. This is further backed up by the underlying hole density distribution $P(x,\tau)$ shown in Fig. \ref{fig.rms_distance_inf_temp}(b) for indicated times. This explicitly shows that the hole remains exponentially localized. I, thus, define the \emph{localization} length as the long-time asymptote of the rms distance. This is plotted in Fig. \ref{fig.rms_distance_inf_temp}(c) for a wide range of spin couplings. In the limit of small spin couplings of $|J| / t\ll 1$, I find very good agreement with a power-law behavior 
\begin{equation}
l_{\rm loc} \to 16 \left[\frac{t}{J}\right]^{2}.
\label{eq.l_loc_inftemp_powerlaw}
\end{equation}
This power-law behavior strongly suggests that the hole will remain localized at any value of $|J|/t$, analogous to the fact that a particle moving in a one-dimensional random potential is localized for any disorder strength, $W$, and only asymptotically move ballistically in the extreme limit of $|J|/t \to 0$. It is worth noting that the scaling behavior is the same as in the usual 1D Anderson model \cite{Fouque1999}. This should be regarded as a non-trivial result for two reasons. Firstly, the disorder potential in this case is correlated between nearest neighbors, since the potential changes at most by $|J|/2$ from site to site. Secondly, the variance of the disorder potential grows linearly with $|x|$ in this model, whereas it is constant in the usual Anderson model. Thirdly, Figure \ref{fig.rms_distance_inf_temp}(c) also reveals that the behavior at any finite temperature is qualitatively different, scaling asymptotically as $t/|J|$. In Sec. \ref{sec.analysis_of_localization}, I will return to these peculiarities.  

\begin{figure}[t!]
\begin{center}
\includegraphics[width=1.0\columnwidth]{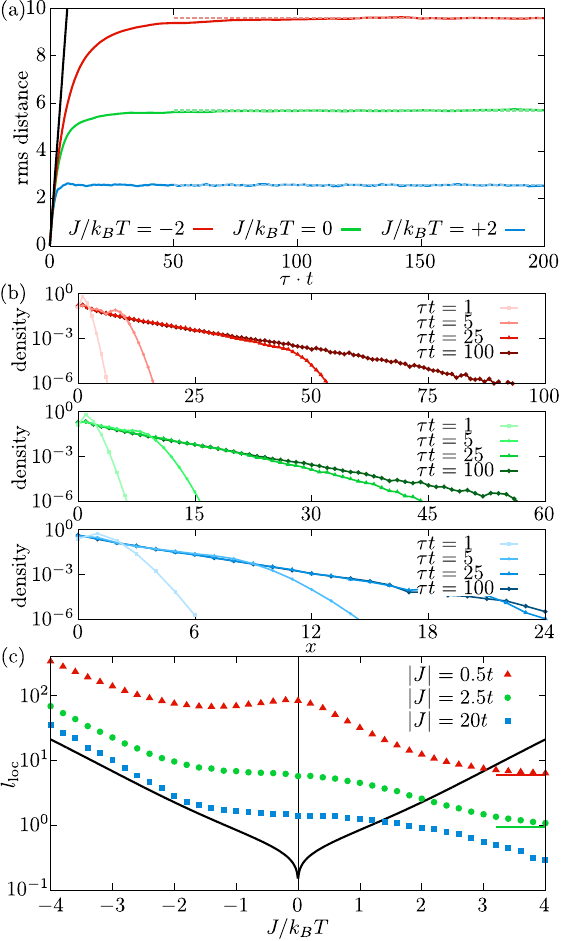}
\end{center}\vspace{-0.5cm}
\caption{(a) Rms distance of the hole versus time for indicated values of the temperature for $|J| / t = 2.5$. The black line again shows ballistic behavior with expansion speed $\sqrt{2}t$, and dashed lines shows the asymptotic flat behavior. (b) Hole density $P(x,\tau)$ at indicated times for the same values of $J / k_BT$ as in (a), again showing exponential localization of the hole. (c) Localization length $l_{\rm loc}$ versus $J / k_BT$ for indicated values of $|J| / t$. On the antiferromagnetic side, $J / k_BT > 0$, $l_{\rm loc}$ decreases and eventually approaches the zero temperature value indicated by the lines to the right. The asymptotic value for $|J| / t = 20$ is below the scale of the plot. On the ferromagnetic side, $J / k_BT < 0$, $l_{\rm loc}$ increases, but \emph{remains finite} for any temperature on a length scale that is larger than the spin-spin correlation length (black line). The estimated statistical errors on the Monte Carlo sampling are smaller than the linewidth/point size and omitted.}
\label{fig.rms_distance_finite_temp} 
\vspace{-0.25cm}
\end{figure} 

\section{Finite temperature behavior} \label{sec.finite_temperature}
For finite temperatures, I employ a standard Metropolis-Hastings Monte Carlo algorithm \cite{Metropolis1953,Hastings1970} to generate a total of $N_\bsigma = 2000$ samples for every investigated value of $k_BT / |J|$. This sampling, in particular, uses single spin flip dynamics. An important question is how to appropriately perform this sampling, as a Monte Carlo algorithm inherently leads to autocorrelation between the samples \cite{Kolafa1986}. This is addressed in Appendix \ref{sec.monte_carlo_sampling}, in which I show that by increasing the sampling interval, i.e. the number of \emph{generated} samples for every \emph{kept} sample the estimated statistical errors dramatically decreases and convergence is observed for sampling intervals above $10^4$ or so, even at very low temperatures. To avoid any sensitivity to this effect, I, therefore, stay well above this threshold and generally use sampling intervals of $10^6$, keeping one in every one million generated samples.

In Fig. \ref{fig.rms_distance_finite_temp}(a), I compare the hereby obtained rms distance dynamics for $|J| / t = 2.5$ at $|J|/k_BT = 2$ to the infinite temperature limit. Although the hole spreads out significantly more on the ferromagnetic side, it \emph{remains localized} at this intermediate temperature. I support this further by showing the hole density distribution in Fig. \ref{fig.rms_distance_finite_temp}(b), which again shows exponential localization of the hole to its origin. In fact, in Fig. \ref{fig.rms_distance_finite_temp}(c) I show the localization length across a broad range of temperatures and values of $|J| / t$, revealing that the hole remains localized for all investigated temperatures and interactions. This shows that the localization phenomenon discovered in the previous section at infinite temperatures is a robust effect and seems to happen as long as the temperature is nonzero. The underlying reason for this robustness, I believe, is that the system, due to its one-dimensional geometry, is always disordered. Therefore, on length scales longer than the spin-spin correlation length $\xi_1(\beta J)$ [see Eq. \eqref{eq.correlation_length}], the hole still sees a randomized potential $V(x)$. To check this intuition, I compare the extracted localization length to the correlation length in Fig. \ref{fig.rms_distance_finite_temp}(c). Indeed, we see that the correlation length follows the trend of localization length on the ferromagnetic side. Moreover, the effect of decreasing temperature is also seen to accelerate when the correlation length starts to exceed $1$. 

\begin{figure}[t!]
\begin{center}
\includegraphics[width=1.0\columnwidth]{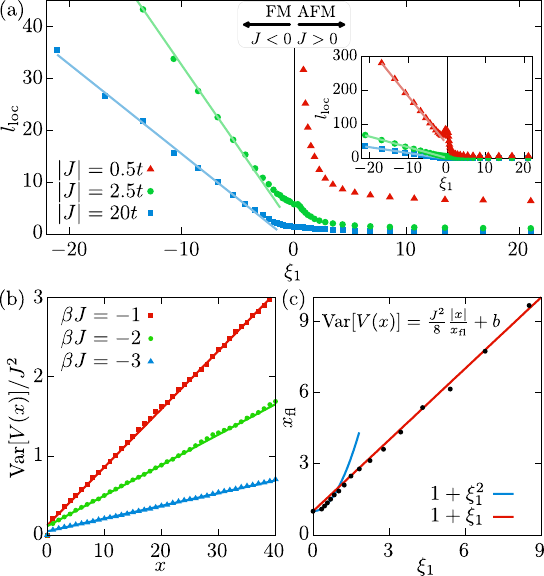}
\end{center}\vspace{-0.5cm}
\caption{(a) Localization length plotted as a function of the spin-spin correlation length $\xi_1(k_B T / J)$ for indicated values of the spin coupling. The inset shows the same plot on a bigger scale. For ferromagnetic couplings (FM, $J < 0$), I multiply $\xi_1$ by $-1$. In this regime, the localization length is linear in $\xi_1$ at low temperatures. For antiferromagnetic couplings (AFM, $J > 0$), the localization length decreases and reaches a plateau at low temperatures as in Fig. \ref{fig.rms_distance_finite_temp}. The localization across all temperatures is traced back to an emergent disorder potential $V(x)$ experiences by the hole, whose variance shown in (b) is linear for all temperatures $T = 1/k_B\beta$. (c) The associated length scale of the potential $x_{\rm fl}$ is plotted as a function of $\xi_1$, showing a linear dependency at large $\xi_1$.}
\label{fig.localization_and_correlation_length} 
\vspace{-0.25cm}
\end{figure}

To get a better understanding of the above effects, I next replot the localization length as a function of the correlation length. This is shown in Fig. \ref{fig.localization_and_correlation_length}(a). This reveals that at low temperatures, corresponding to $\xi_1(\beta J) \gg 1$, the localization length becomes linear in the correlation length for ferromagnetic couplings,
\begin{align}
l_{\rm loc}(\beta J) = \gamma \times \xi_1(\beta J).
\label{eq.loc_length_vs_corr_length}
\end{align}
The analysis additionally unveils that the prefactor $\gamma$ \emph{increases with decreasing} $|J| / t$. In this manner, the hole motion only delocalizes in the asymptotic limit of zero temperature. Here, all spins align at $T = 0$ and the magnetic potential obtained in Eqs. \eqref{eq.intraleg_potential} and \eqref{eq.transleg_potential} vanishes identically, whereby the hole is free to move ballistically through the system. 

On the antiferromagnetic side, the localization length is conversely seen to decrease. The reason is that at zero temperature, the accompanying magnetic potential defined in Eqs. \eqref{eq.intraleg_potential} and \eqref{eq.transleg_potential} increases linearly with distance, $V(x) = J / 2 [|x| + 1]$, as also obtained previously \cite{Nielsen2023_1}, which localizes the hole more strongly than in the disordered case. Moreover, the decrease in localization length is seen to be very rapid at low $\xi_1$, but quickly saturates as $\xi_1 \gg l_{\rm loc}$. This is also intuitively clear, since the correlation length sets the typical length scale over the system is ordered. Therefore, if the localization length is much smaller than the correlation length, it does not see the long-range disorder. 

\section{Semi-classical analysis of localization} \label{sec.analysis_of_localization}
To qualitatively understand the dependency on spin coupling and temperature seen in the previous two sections, I analyze these dependencies using a semi-classical energy argument. First, For large values of $|J|/t$ and for ferromagnetic spin couplings, $J < 0$, the hole meets a potential wall of order $\sim\!|J|$ that it cannot pass with its minuscule kinetic energy $\sim\!t$, as soon as its effective potential $V_\bsigma(x)$ deviates from zero. The length scale for this to happen is set by the size of ferromagnetic clusters, which in turn is given by the spin-spin correlation length $\xi_1$ [Eq. \eqref{eq.correlation_length}]. This explains the low-temperature behavior shown in Fig. \ref{fig.localization_and_correlation_length}(a) in the limiting case of $|J| \gg t$ in terms of an immediate backscattering of the hole as soon as the potential changes away from $0$.

Second, keeping this backscattering in mind, for intermediate to low values of $|J| / t$, we may instead ask at what length scale the initial kinetic energy will \emph{typically} match the potential energy. To understand this, we first have to realize that the effective potential both has a positive mean value, $\braket{V_{\bsigma,\perp}(x)} > 0$, as a well as nonzero fluctuations, ${\rm Var}[V_{\bsigma}(x)] > 0$, given by
\begin{align}
\braket{V_{\bsigma}(x)} &= \frac{|J|}{2} \frac{|x|}{x_{\rm ave}} + b_{\rm ave}, \nn \\
{\rm Var}[V_{\bsigma}(x)] &= \frac{J^2}{8} \frac{|x|}{x_{\rm fl}} + b_{\rm fl}, 
\label{eq.potential_bias_and_variance}
\end{align}
both of which scale linearly in the distance $|x|$. The mean value is calculated analytically from Eq. \eqref{eq.transleg_potential}, leading to the length scale 
\begin{align}
x_{\rm ave} = \frac{2}{C(1) - C(\sqrt{2})},  
\end{align}
defining the nearest and next-nearest neighbor correlators $C(1) = 4\braket{\sigma_{1,0}\sigma_{2,0}}$, $C(\sqrt{2}) = 4\braket{\sigma_{1,0}\sigma_{2,1}}$ across the ladder. Importantly, $x_{\rm ave}$ scales as $\exp(3\beta|J|/2) \propto \xi_1^{3/2}$ at low temperatures. 

Moreover, the linearity of the variance is found not only to be true at infinite temperatures [Eq. \eqref{eq.variance_potential} with $x_{\rm fl} = 1$], but also at any finite temperature. This is establishes numerically in Fig. \ref{fig.localization_and_correlation_length}(b), and the temperature dependent length scale is found to be closely tied to the spin-spin correlation length [Fig. \ref{fig.localization_and_correlation_length}(c)]
\begin{align}
\begin{matrix*}[l]
x_{\rm fl}(\beta J) = 1 + [\xi_1(\beta J)]^2, & \xi_1(\beta J) \ll 1, \\
x_{\rm fl}(\beta J) = 1 + \xi_1(\beta J), & \xi_1(\beta J) \gg 1.
\end{matrix*}
\end{align}
The crossover between the two behaviors is very rapid and happens around $\xi_1(\beta J) = 1$, corresponding to $k_BT \simeq -J$. Importantly, this shows a \emph{linear} dependency on $\xi_1$ at low temperatures. A physically intuitive way to understand this is to imagine a typical state at very low temperatures in the ladder. This will consists of domains of size $\xi_1$. As a result, every time the hole has traversed a distance of $\xi_1$ it will \emph{at random} go up or down by $|J|/2$, performing a classical random walk with length scale $x_{\rm fl} \simeq \xi_1$ instead of $x_{\rm fl} = 1$ in Eq. \eqref{eq.variance_potential}. 

We are now ready to estimate the localization length. First, if the bias of the potential dominates over its fluctuations at the relevant length scale, $\braket{V_{\bsigma}} > ({\rm Var}[V_{\bsigma}] )^{1/2}$, then we may simply equate the initial kinetic energy to the bias: $t = \braket{V_{\bsigma}(l_{\rm ave})}$. This gives the length scale
\begin{align}
l_{\rm ave} = \frac{4}{C(1) - C(\sqrt{2})} \frac{t}{|J|},
\label{eq.localization_length_from_mean_potential}
\end{align}
Secondly, if the fluctuations of the potential dominate $({\rm Var}[V_{\bsigma}] )^{1/2} > \braket{V_{\bsigma}}$, the backscattering happens on a length scale set by $t = ({\rm Var}[V_{\bsigma}] )^{1/2}$. This gives a fluctuation induced localization length scale
\begin{align}
l_{\rm fl} = 8 x_{\rm fl}(\beta J) \left[\frac{t}{J}\right]^2.
\label{eq.localization_length_from_variance}
\end{align}
From Eqs. \eqref{eq.localization_length_from_mean_potential} and \eqref{eq.localization_length_from_variance}, we are now ready to understand the intricate dependencies of the localization on spin coupling and temperature. At infinite temperatures, the correlators $C(1)$ and $C(\sqrt{2})$ both vanish and the localization length set by the average potential in Eq. \eqref{eq.localization_length_from_mean_potential} diverges. Moreover, in this limit $x_{\rm fl}(\beta J) \to 1$, and so the localization becomes solely determined by the fluctuations given in Eq. \eqref{eq.localization_length_from_variance} and goes as $l_{\rm fl} \sim [t/J]^2$ in agreement with Eq. \eqref{eq.l_loc_inftemp_powerlaw} and Fig. \ref{fig.rms_distance_inf_temp}(c). For any \emph{fixed finite temperature} and decreasing values of $|J| / t$, on the other hand, the fluctuation length scale $l_{\rm fl}$ will eventually surpass the bias length scale $l_{\rm ave}$. As a result, the localization will eventually be set by the bias following Eq. \eqref{eq.localization_length_from_mean_potential}, as evident in Fig. \ref{fig.rms_distance_inf_temp}(c). 

Finally, the bias and fluctuation length scales in the low-temperature limit goes as $l_{\rm ave} \sim \xi_1^{3/2} t/|J|$, and $l_{\rm fl} \sim [1 + \xi_1] [t/J]^2$. Therefore, at fixed spin couplings $|J|/t$ and decreasing temperatures, $\beta J \to -\infty$, eventually the fluctuation length scale becomes the shortest, $l_{\rm fl} < l_{\rm ave}$, and the localization length is once again set by the fluctuations. This explains the simple proportionality with the correlation length $\xi_1$ observed not only at strong spin couplings $|J|/t \gg 1$, but also at intermediate and weak values. 

Moreover, this competition of two effects, localization due to the fluctuations and due to a biased potential, also explains the non-monotonic dependency of the localization length on the temperature at a low value of $|J|/t = 0.5$ seen in Figs. \ref{fig.rms_distance_finite_temp}(c) and \ref{fig.localization_and_correlation_length}(a) in the following sense. At infinite temperatures, the localization length scales as $(t/J)^2$, but as soon as the temperature drops, the bias of the potential becomes nonzero, and the localization length now scales as $t/|J|$, hereby contracting the hole cloud. This leads to a drop in $l_{\rm loc}$ until temperatures are so low that the bias decreases again and the localization length scales as $\xi_1 \times (t/J)^2$.

\section{Detection in optical lattices with Rydberg-dressed atoms} \label{sec.Rydberg_experiment}
In this section, I describe how the discovered localization phenomenon can be detected using current experimental setups with Rydberg-dressed atoms \cite{Zeiher2016}. Such a setup natively implements finite range density-density interactions,
\begin{equation}
\Ham_J = \frac{1}{2}\sum_{\bi \neq \bj} J_{\bi\bj} \hat{n}_{\bi\uparrow}\hat{n}_{\bj\uparrow},
\label{eq.H_J_rydberg}
\end{equation}
of the internal atomic state $\ket{\uparrow}$ that is being dressed by a higher-lying Rydberg state via an optical light field. Here, $J_{\bi\bj} = J_0 / (1 + (|\bi - \bj| / r_c)^6)$ takes on a soft-core shape, with $r_c$ the soft-core size \cite{Henkel2010}. The $\ket{\downarrow}$ state remains uncoupled from the light field and does not experience the interaction. Furthermore, an interstate Feshbach resonance may be used to drive the system into the Mott-insulating phase, such that there is at most a single spin on each site. Crucially important, the associated onsite interaction $U$ between $\ket{\downarrow}$ and $\ket{\uparrow}$ can be increased independently of the light-induced interaction $J_{\bi\bj}$. As a result, low-energy spin-exchange interactions $\propto 4t^2 / U$ \cite{Dagotto1994} can be made negligible compared to the interactions of Eq. \eqref{eq.H_J_rydberg} on the investigated timescales. 

The density-density interaction in Eq. \eqref{eq.H_J_rydberg} can equivalently be thought of as  asymmetric finite-range Ising interactions. Doping the system with holes and allowing the spins to tunnel along the ladder with rate $t$, hereby, realizes a modified Ising $t$-$J$ model that can be used to test the predictions made in this Article. In particular, we can express Eq. \eqref{eq.H_J_rydberg} in terms of spin excitation and hole operators as
\begin{equation}
\Ham_J = \frac{1}{2}\sum_{\bi \neq \bj} \!J_{\bi\bj} \big[1 \!-\! \hat{s}^\dagger_\bi\hat{s}_\bi\big]\big[1 \!-\! \hat{s}^\dagger_\bj\hat{s}_\bj\big]\big[1 \!-\! \hat{h}^\dagger_\bi\hat{h}_\bi\big]\big[1 \!-\! \hat{h}^\dagger_\bj\hat{h}_\bj\big],
\label{eq.H_J_rydberg_spins_and_holes}
\end{equation}
and I will then analyze the motion of a hole starting out at $\bi = {\bf 0}$. While precise experimental control of the temperature is generally difficult, we can take an alternative route to investigate the propagation of the hole in an effectively disordered medium. In particular, the system can be initialized with a hole at $\bi = {\bf 0}$ by applying a strong repulsive light field to that site \cite{Ji2021}. Moreover, I assume that the spins are initially all polarized into the non-interacting $\ket{\downarrow}$ state, such that $\ket{\Psi_{\pi / 2}} = \prod_{\bi\neq {\bf 0}} \hat{c}^\dagger_{\bi\downarrow} \ket{0}$. Then, a depolarizing field can be applied to mix the $\ket{\uparrow}$ and $\ket{\downarrow}$ states on each site with a specified mixing angle $\theta$
\begin{align}
\ket{\Psi_\theta} = \prod_{\bi \neq {\bf 0}} \left[\cos(\theta)\hat{c}^\dagger_{\bi\uparrow} + \sin(\theta)\hat{c}^\dagger_{\bi\downarrow}\right]\ket{0}. 
\label{eq.psi_theta}
\end{align}
With this as the initial state for a given mixing angle $\theta$, the light field on site $\bi = {\bf 0}$ can be turned off such that the hole is now allowed to tunnel \emph{along the ladder}, as described by $\Ham_t$ in Eq. \eqref{eq.H_t_holstein_primakoff}. The ability to turn off hopping between the legs relies on an additional energy offset between the legs \cite{Hirthe2023}. Although this at face value is different from the nonzero temperatures considered previously in the Article, the dynamics of the hole can be described in a completely equivalent manner. In particular, the probability of finding the hole at site $x$ at time $\tau$
\begin{align}
P(x,\tau) &= \bra{\Psi_\theta}e^{+i\Ham \tau}\hat{h}^\dagger_{1,x} \hat{h}_{1,x} e^{-i\Ham \tau}\ket{\Psi_\theta} \nn \\
&= \sum_{\bsigma} p_{\bsigma}(\theta) |C_\bsigma(x,\tau)|^2,
\label{eq.prob_hole_dynamics_rydberg}
\end{align}
takes on exactly the same form as Eq. \eqref{eq.prob_hole_dynamics} for the nonzero temperature case. The probabilities $p_{\bsigma}(\theta)$ are now, however, not given by the thermal statistics, but a binomial distribution depending on the number of spin-$\downarrow$ atoms, $N_\downarrow(\bsigma)$ in the spin realization $\bsigma$
\begin{align}
p_{\bsigma}(\theta) = [\sin^2 \theta ]^{N_\downarrow(\bsigma)} [\cos^2 \theta ]^{N - N_\downarrow(\bsigma)}. 
\label{eq.spin_distribution_rydbergs}
\end{align}
As a result, such an experimental setup simulates the thermally induced localization phenomenon described in the previous sections. Here, the ferromagnetic states correspond to $\theta = 0,\pi/2$, whereas the infinite temperature limit corresponds to $\theta = \pi / 4$. Note that this defines another way of effectively achieving an initial infinite-temperature thermal ensemble. For other value of $\theta$, there is strictly speaking no one-to-one correspondence with a specific temperature, but the variation of $\theta$ in the interval $[0,\pi / 2]$ qualitatively describes the same behavior as a varying temperature. Moreover, as the hole hops through the system, it experiences a magnetic potential akin to Eqs. \eqref{eq.intraleg_potential} and \eqref{eq.transleg_potential}. Specifically, for a given initial spin realization $\bsigma$, a certain subset of the sites $S_\uparrow(\bsigma,0)$ contains spin-$\uparrow$ atoms. This leads to the overall energy offset
\begin{align}
V_0 = \frac{1}{2}\sum_{\bi,\bj\in S_\uparrow(\bsigma,0)} J_{\bi\bj}.
\label{eq.energy_offset_rydbergs}
\end{align}
As the hole hops, the surpassed spin hops one site in the opposite direction. As a result, the subset of sites $S_\uparrow(\bsigma,x)$ with spin-$\uparrow$ depends on the position of the hole $x$. The resulting magnetic potential is then simply the magnetic energy differences 
\begin{align}
V_\bsigma(x) = \frac{1}{2}\sum_{\bi,\bj\in S_\uparrow(\bsigma,x)} J_{\bi\bj} - V_0.
\label{eq.potential_rydbergs}
\end{align}
experienced as the hole moves through the system. With this at hand, the computation of the hole dynamics now follows the same recipe as in Sec. \ref{sec.finite_temperature}. In this case, I assume a finite size of the system with hard-wall boundary conditions and a total length $N = 41$ to properly describe a feasible experimental setup. The Metropolis-Hastings algorithm again uses $N_\bsigma = 2000$ samples, and is benchmarked by comparing the achieved magnetization per spin to the exact value of $[\cos^2 \theta - \sin^2 \theta]/2$. I find agreement within $1\%$ for any value of $\theta$. 

\begin{figure}[t!]
\begin{center}
\includegraphics[width=1.0\columnwidth]{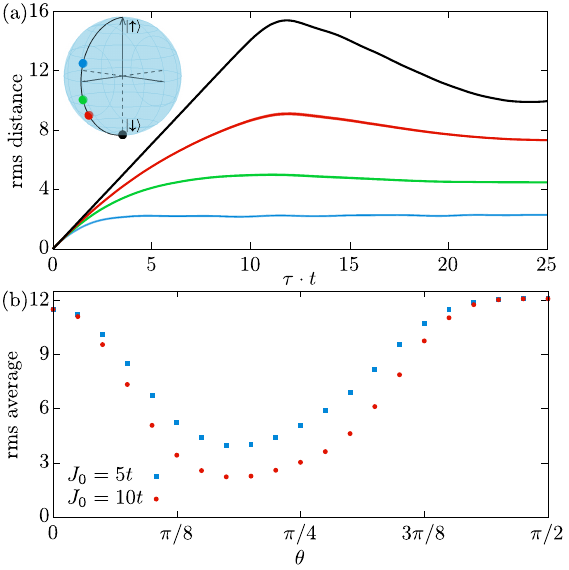}
\end{center}\vspace{-0.5cm}
\caption{(a) Rms distance versus time for $J_0 = 10t$ and indicated mixing angles shown on the Bloch sphere (inset), corresponding to $\theta = \pi/2,0.7\pi/2,0.6\pi/2,0.4\pi/2$ for black, red, green and blue respectively. For $\theta = \pi/2$ (black), the hole is free to propagate and only attains an oscillatory behavior due to the finite size of the system (total length of $N = 41$). (b) Long-time average of rms distance for indicated values of $J_0$ as a function of the mixing angle $\theta$, following the black line on the Bloch sphere in (a). This shows a minimum between $\theta = \pi/8$ and $\theta = \pi / 4$ due to localization. The estimated standard errors are smaller than the dot size and are omitted for clarity. I use a soft-core size of $r_c = 1$.}
\label{fig.rydberg_dynamics} 
\vspace{-0.25cm}
\end{figure} 

Figure \ref{fig.rydberg_dynamics}(a) shows the resulting dynamics of the rms distance for indicated points on the Bloch sphere. Here, the north and south poles correspond to all spins pointing up and down, respectively, such that the mixing angle $\theta$ is nothing but half the polar angle on the Bloch sphere. The dynamics is qualitatively similar to the cases shown in Figs. \ref{fig.rms_distance_inf_temp}(a) and \ref{fig.rms_distance_finite_temp}(a). The only major difference is that the free motion of the hole at $\theta = \pi / 2$ now becomes oscillatory due to the finite size of the system. We see that as the mixing angle goes away from $\theta = \pi/2$, the hole starts to localize. This is shown in more detail in Fig. \ref{fig.rydberg_dynamics}(b), where the long-time average of the rms distance is plotted as a function of the mixing angle. At $\theta = 0,\pi/2$, the average rms distance of the hole settles around half the distance to the edge of the system. However, as the $50$-$50$ spin mixing at $\theta = \pi / 4$ is approached, this dramatically decreases and reaches a minimum around $\theta = 0.75 \pi / 4$. This is a direct signature of localization of the hole. Indeed, the observed localization length around the minimum is no longer sensitive to the system size. One may reasonably wonder why the minimum is not located exactly at $\theta = \pi / 4$. The reason is that the spin interactions in Eq. \eqref{eq.H_J_rydberg} are not symmetric in spin-$\uparrow$ and -$\downarrow$, and indeed vanishes identically for the latter states. Moreover, the limiting values at the top and bottom of the Bloch sphere, $\theta = 0,\pi/2$ respectively, do not perfectly coincide. This is because the repulsive interactions of the spin-$\uparrow$ atoms in the case of $\theta = 0$ favors the hole not to move all the way out to the edge of the system. This is a very minor effect that only shows up at the sites just before the edge. 

Crucially, this analysis directly shows that the localization can be probed on reasonably short timescales of just $\tau = 5/t$. This is highly important for the considered experimental protocol, because the Rydberg-dressed spin-$\uparrow$ atoms inherit some of the decay of the high-lying Rydberg state. Here, it is also beneficial that the localization can be probed on the side where there is a majority of spin-$\downarrow$ atoms ($\pi / 4 < \theta < \pi / 2$), making this inherent decay less severe. This analysis, thus, establishes that the thermally induced localization discovered in the present Article may be realistically probed in current experimental platforms using Rydberg-dressed atoms. 

Moreover, I emphasize that this phenomenon should show up in any system that has polarized interactions, like the Ising cases considered here, and short-range hopping of a dopant. This suggests that one could also come up with a protocol using dipolar gases in optical lattices \cite{Trefzger2011,Chomaz2023} or trapped ions \cite{Kranzl2023}, in which a similar localization could happen.  

\section{Conclusions and outlook} \label{sec.conclusions}
In this Article, I have described a novel localization phenomenon of dopants in Ising-type magnetic spin ladders. The effect arises \emph{not} due to inherent disorder in the system Hamiltonian, but as an \emph{emergent} phenomenon \cite{DeRoeck2014} due to thermal spin fluctuations. In particular, since the system is one-dimensional, it is disordered at any nonzero temperature. Therefore, even for ferromagnetic spin couplings for which one might expect the hole to delocalize completely, I show that it \emph{remains} localized across a huge range of spin couplings $J / t$ and temperatures $k_B T / J$. The effect is traced back to a \emph{disorder potential} experienced by the hole, whose strength increases towards infinite temperatures. In this infinite temperature limit, the dynamics no longer depends on the sign of the spin interactions and becomes a universal function of $|J|/t$. Moreover, I showcased have the localization phenomenon may be explored using current experimental platforms with Rydberg-dressed atoms in optical lattices. 

The results strongly suggest that in general Ising environments in the disordered phase, dopants moving in one spatial direction will remain localized. As a result, the system will be an insulator provided that the inter-dopant spacing is large compared to the localization length. This suggests the possibility of a \emph{reversed} metal-insulator transition -- or at least crossover -- around the Curie temperature $T_C$, such that dopants are localized \emph{above} $T_C$, and delocalizes \emph{below} $T_C$. While this transition occurs at zero temperature, $T_C = 0$, in the current one-dimensional setup, it would be interesting to study the same scenario in two and three spatial dimensions, in which the Curie temperature is nonzero. 

Additionally, it is important to analyze the robustness of the localization phenomenon going away from the idealized models considered in the present Article. For example, does the same phenomenology arise when the dopants are allowed to move in two or three dimensions? Here, studies of dopant motion in a \emph{non-interacting} two-dimensional spin lattice at infinite temperatures \cite{Carlstrom2016,Nagy2017} shows that there are crucial qualitative differences. In particular, these results suggest diffusive motion of the dopant due to lack of path interferences in the disordered medium, which is in contrast to the ballistic behavior obtained for the one-dimensional motion in the present setup in this limit of $J/t \to 0$. Turning on spin interactions in such a two-dimensional setup \cite{Hahn2022}, and carefully analyzing the long-time dynamics for a broad range of spin interactions could help to answer this question. Along the same lines, one could also analyze what happens in the presence of spin-exchange processes, whose analysis should be amenable to matrix product states approaches. Na{\"i}vely, the presented results suggest that the spin and charge degrees of freedom become bound in such a disordered phase, only allowing the hole to move with the slow co-propagation of the trailing spin. For this reason, there could be very intriguing behaviors in a similar two-leg ladder setup in e.g. an XXZ model \cite{White2015,Zhu2015}, as one tunes the anisotropy of the spin couplings towards the isotropic Heisenberg model. In a similar spirit, it would also be interesting to investigate what happens in the presence of an external heat bath that drives the spins back to thermal equilibrium. One could imagine that the hole can now diffuse through the system depending on the coupling with the external bath. Finally, it would be intriguing to understand whether two dopants can actually co-propagate in the ladder, establishing a novel pairing by disorder mechanism. 

\begin{acknowledgments}
The author thanks Pascal Weckesser, Johannes Zeiher, Marton Kanasz-Nagy, Pavel Kos, Dominik Wild, and J. Ignacio Cirac for valuable discussions. A special thanks to Pascal Weckesser for providing valuable references. This Article was supported by the Carlsberg Foundation through a Carlsberg Internationalisation Fellowship, grant number CF21\_0410. 
\end{acknowledgments}

\appendix

\section{Thermodynamics in the absence of dopants} \label{app.thermodynamics}
The thermodynamics of the system at half filling can be studied using a transfer matrix technique closely related to the analysis of a single chain. The Hamiltonian of the system is simply the nearest neighbor Ising Hamiltonian
\begin{align}
\Ham_J = J \sum_{\braket{\bi,\bj}} \hat{S}^{(z)}_\bi\hat{S}^{(z)}_\bj \to |J| \sum_{\braket{\bi,\bj}} \hat{S}^{(z)}_\bi\hat{S}^{(z)}_\bj.
\end{align}
In the last expression, I perform a local rotation on every second site $\hat{S}^{(z)}_\bj \to -\hat{S}^{(z)}_\bj$ for ferromagnetic couplings, $J < 0$. This shows that the thermodynamics is equivalent for antiferro- and ferromagnetic couplings. Denoting the spin configurations in legs 1 and 2 respectively $\bsigma_1$ and $\bsigma_2$, we get the partition function in the canonical ensemble
\begin{align}
Z_J &= \tr[e^{-\beta \Ham_J}] \nn \\
&= \sum_{\bsigma_1,\bsigma_2} e^{\beta |J|\sigma_{1,1}\sigma_{1,2}} e^{\beta |J|\sigma_{2,1}\sigma_{2,2}} e^{\beta |J| \sigma_{1,1} \sigma_{2,1}} e^{\beta |J| \sigma_{1,2} \sigma_{2,2}} \times \nonumber \\
&\dots \times e^{\beta |J|\sigma_{1,N}\sigma_{1,1}} e^{\beta |J|\sigma_{2,N}\sigma_{2,1}}
\end{align} 
for a system of length $N$ with periodic boundary conditions. Defining the $4\times 4$ transfer matrix
\begin{align}
\!\!\!\!V_{\sigma_{1,1},\sigma_{1,2}}^{\sigma_{2,1},\sigma_{2,2}} =\, & e^{\beta |J|\left[\sigma_{1,1}\sigma_{1,2} + \sigma_{2,1}\sigma_{2,2} + \sigma_{1,1} \sigma_{2,1} / 2 + \sigma_{1,2} \sigma_{2,2} / 2\right]},\!\!\!
\end{align}
we can then write the partition function much more concisely as
\begin{align}
Z_J &= \tr[e^{-\beta \hat{H}_J}] = \sum_{\bsigma_1,\bsigma_2} V_{\sigma_{1,1},\sigma_{1,2}}^{\sigma_{2,1},\sigma_{2,2}} \times \dots \times V_{\sigma_{1,N},\sigma_{1,1}}^{\sigma_{2,N},\sigma_{2,1}} \nn \\
&= \sum_{\{\eta_l\}_{l=1}^N} V_{\eta_1,\eta_2}V_{\eta_2,\eta_3} \times \dots \times V_{\eta_N,\eta_1} = \tr[V^N]. 
\end{align} 
In the second line, I used the states $\ket{\sigma_1,\sigma_2}$ in the ordered basis $
\{\ket{\uparrow\uparrow},\ket{\downarrow\uparrow},\ket{\uparrow\downarrow},\ket{\downarrow\downarrow}\}$, such that $\eta = 1,2,3,4$ refers to these elements respectively. The rows and columns of $V$ correspond to different values of $(\sigma_{1,1},\sigma_{2,1})$ and $(\sigma_{1,2},\sigma_{2,2})$, respectively. Hence,
\begin{equation}
V = \begin{bmatrix} 
e^{+3\beta|J|/4} & 1 & 1 & e^{-\beta|J|/4} \\ 
1 & e^{+\beta|J|/4} & e^{-3\beta|J|/4} & 1 \\
1 & e^{-3\beta|J|/4} & e^{+\beta|J|/4} & 1 \\
e^{-\beta|J|/4} & 1 & 1 & e^{+3\beta|J|/4}
\end{bmatrix}.
\end{equation}
The problem has now been reduced to finding the $4$ eigenvalues, $v_1,\dots,v_4$, of the transfer matrix $V$. In fact, letting $v_1$ be the largest eigenvalue, we get
\begin{align}
Z_J &= \tr[V^N] = \sum_{j} \bra{v_j} V^N \ket{v_j} = \sum_j v_j^N \to v_1^N, 
\label{eq.partition_function_v1}
\end{align}
as $N\to \infty$. So we only need the largest eigenvalue $v_1$. From here, the free energy per spin is (there are $2N$ spins) 
\begin{align}
F_0 = -\frac{1}{2\beta N} \ln Z_J = -\frac{1}{2\beta}\ln v_1. 
\label{eq.free_energy_v1}
\end{align}
Diagonalizing a $4\times4$ matrix is not trivial, however, since it in general means that we have to solve a fourth order characteristic polynomial. However, we may use that the Hamiltonian does not couple the triplet $\{\ket{\uparrow\uparrow},\ket{\downarrow\downarrow}, (\ket{\uparrow\downarrow} + \ket{\downarrow\uparrow}) / \sqrt{2}\}$ and singlet $\{(\ket{\uparrow\downarrow} - \ket{\downarrow\uparrow}) / \sqrt{2}\}$ subspaces. Transforming from the former to the latter basis is done via
\begin{align}
U = \begin{bmatrix} 1 & 0 & 0 & 0 \\ 0 & 0 & 2^{-1/2} & 2^{-1/2} \\ 0 & 0 & 2^{-1/2} & -2^{-1/2} \\ 0 & 1 & 0 & 0 \end{bmatrix}.
\end{align}
Expressing the transfer matrix in the triplet-singlet basis yields 
\begin{widetext}
\begin{align}
\tilde{V} &= U^\dagger V U = \begin{bmatrix} 
e^{+3\beta|J|/4} & e^{-\beta|J|/4} & \sqrt{2} & 0  \\ 
e^{-\beta|J|/4} & e^{+3\beta|J|/4} & \sqrt{2} & 0 \\
\sqrt{2} & \sqrt{2} & e^{+\beta|J|/4} + e^{-3\beta|J|/4} & 0 \\
0 & 0 & 0 & e^{+\beta|J|/4} - e^{-3\beta|J|/4}.
\end{bmatrix}
\end{align}
\end{widetext}
Diagonalizing the remaining $3\times 3$ matrix, the eigenvectors are
\begin{align}
\ket{v_i} &= \frac{1}{\sqrt{A_i}}\begin{bmatrix} \frac{v_i - (e^{\beta|J|/4} + e^{-3\beta|J|/4})}{2\sqrt{2}} \\ \frac{v_i - (e^{\beta|J|/4} + e^{-3\beta|J|/4})}{2\sqrt{2}} \\ 1 \\ 0 \end{bmatrix},\; i = 1,2 \nn \\
\ket{v_3} &= \frac{1}{\sqrt{2}}\begin{bmatrix} 1 \\ -1 \\ 0 \\ 0 \end{bmatrix}, \ket{v_4} = \begin{bmatrix} 0 \\ 0 \\ 0 \\ 1\end{bmatrix},
\end{align}
with $A_i = ([v_i - (e^{\beta|J|/4} + e^{-3\beta|J|/4})]^2 + 4)/4$. The eigenvalues are
\begin{align}
\!\!\!v_1 =&\; 2\cosh \frac{1}{2}\beta |J| \, \cosh \frac{1}{4}\beta |J|  \nn \\
&+ 2\sqrt{\left(\cosh \frac{1}{2}\beta |J| \, \cosh \frac{1}{4}\beta |J|\right)^2 - \sinh^2 \frac{1}{2}\beta|J|}, \nn \\
\!\!\!v_2 =&\; 2\cosh \frac{1}{2}\beta |J| \, \cosh \frac{1}{4}\beta |J|  \nn \\
&- 2\sqrt{\left(\cosh \frac{1}{2}\beta |J| \, \cosh \frac{1}{4}\beta |J|\right)^2 - \sinh^2 \frac{1}{2}\beta|J|}, \nn \\
\!\!\!v_3 =&\; 2e^{\beta|J|/4}\sinh \frac{1}{2}\beta|J|, v_4 = 2e^{-\beta|J|/4}\sinh \frac{1}{2}\beta|J|.\!
\end{align}
I find that $v_1$ is the largest eigenvalue for any value of $\beta J$. The free energy of the system $F_0 = -\frac{1}{2\beta}\ln v_1$, hereby, correctly approaches the ground state energy $-3|J| / 8$, at zero temperature: $\beta |J|\to \infty$. Finally, we need the spin-spin correlation function 
\begin{equation}
C_{z}(d) = 4\braket{\hat{S}^{(z)}_{1,1}\hat{S}^{(z)}_{1,1+d}}
\end{equation}
to compare with the localization length of the hole. I use a similar method to the above to find an analytic solution. First, note that
\begin{align}
Z_J C_{z}(d) &= \tr\left[4\hat{S}^{(z)}_{1,1}\hat{S}^{(z)}_{1,1+d}e^{-\beta \Ham_J}\right] \nn \\
&= \tr\left[4\hat{S}^{(z)}_{1,1}\prod_{j=1}^{d-1}4(\hat{S}^{(z)}_{1,1+j})^2\hat{S}^{(z)}_{1,1+d}e^{-\beta \Ham_J}\right] \nn \\
&= \tr\left[\prod_{j=1}^{d}(4\hat{S}^{(z)}_{1,j}\hat{S}^{(z)}_{1,j+1})e^{-\beta \Ham_J}\right].
\end{align}
Here, I use that $(\hat{S}^{(z)}_{1,1+j})^2 = 1/4$ for all the Ising eigenstates. Expressing the above equation in terms of these eigenstates $\ket{\bsigma_1,\bsigma_2}$, thus, yields
\begin{align}
Z C_{z}(d) = &\sum_{\bsigma_1,\bsigma_2} (4\sigma_{1,1}\sigma_{1,2}) V_{\sigma_{1,1},\sigma_{1,2}}^{\sigma_{2,1},\sigma_{2,2}} \times \dots \nn \\
&\times (4\sigma_{1,d}\sigma_{1,d+1}) V_{\sigma_{1,d},\sigma_{1,d+1}}^{\sigma_{2,d},\sigma_{2,d+1}} \times V_{\sigma_{1,d+1},\sigma_{1,d+2}}^{\sigma_{2,d+1},\sigma_{2,d+2}}\times \nn \\
& \dots \times V_{\sigma_{1,N},\sigma_{1,1}}^{\sigma_{2,N},\sigma_{2,1}} = \tr[C^{d}V^{N-d}]
\end{align}
In the last equality, I let $C_{\sigma_{1,1},\sigma_{1,2}}^{\sigma_{2,1},\sigma_{2,2}} = (4\sigma_{1,1}\sigma_{1,2}) V_{\sigma_{1,1},\sigma_{1,2}}^{\sigma_{2,1},\sigma_{2,2}}$. The correlator matrix 
\begin{equation}
C = \begin{bmatrix} 
e^{+3\beta|J|/4} & -1 & 1 & -e^{-\beta|J|/4} \\ 
-1 & e^{+\beta|J|/4} & -e^{-3\beta|J|/4} & 1 \\
1 & -e^{-3\beta|J|/4} & e^{+\beta|J|/4} & -1 \\
-e^{-\beta|J|/4} & 1 & -1 & e^{+3\beta|J|/4}
\end{bmatrix}.
\end{equation}
is very similar to the transfer matrix, and simply attains sign flip with respect to $V$, whenever $\sigma_{1,1}$ and $\sigma_{1,2}$ differ in sign. I also transform this matrix to the singlet-triplet basis 
\begin{widetext}
\begin{align}
\tilde{C} &= U^\dagger C U = \begin{bmatrix} 
e^{+3\beta|J|/4} & -e^{-\beta|J|/4} & 0 & -\sqrt{2}  \\ 
-e^{-\beta|J|/4} & e^{+3\beta|J|/4} & 0 & \sqrt{2} \\
0 & 0 & e^{+\beta|J|/4} - e^{-3\beta|J|/4} & 0 \\
-\sqrt{2} & \sqrt{2} & 0 & e^{+\beta|J|/4} + e^{-3\beta|J|/4}
\end{bmatrix}
\end{align}
\end{widetext}
The eigenvectors of $\tilde{C}$ are closely tied to those of $\tilde{V}$. I get
\begin{align}
\ket{c_i} &= \frac{1}{\sqrt{A_i}}\begin{bmatrix} \frac{v_i - (e^{\beta|J|/4} + e^{-3\beta|J|/4})}{2\sqrt{2}} \\ -\frac{v_i - (e^{\beta|J|/4} + e^{-3\beta|J|/4})}{2\sqrt{2}} \\ 0 \\ -1 \end{bmatrix},\; i = 1,2 
\end{align}
\begin{align}
\ket{c_3} &= \frac{1}{\sqrt{2}}\begin{bmatrix} 1 \\ 1 \\ 0 \\ 0 \end{bmatrix}, \ket{c_4} = \begin{bmatrix} 0 \\ 0 \\ 1 \\ 0\end{bmatrix}.
\end{align}
The corresponding eigenvalues are the same as for the transfer matrix: $c_i = v_i$ for $i = 1,2,3,4$. I am now ready to calculate the spin-spin correlation function. I get
\begin{align}
\!\!\!&Z_J C_{z}(d) = \tr[C^{d}V^{N-d}] = \tr[\tilde{C}^{d}\tilde{V}^{N-d}] \nn \\
\!\!\!&= \sum_i \bra{v_i}\tilde{C}^{d}\tilde{V}^{N-d}\ket{v_i} = \sum_{i,j} \bra{v_i}\tilde{C}^{d}\ket{c_j}\bra{c_j}\tilde{V}^{N-d}\ket{v_i} \nn \\
\!\!\!&= \sum_{i,j} c_j^d v_i^{N - d} |\braket{v_i|c_j}|^2 \to v_1^{N - d}\sum_{j} c_j^d |\braket{v_1|c_j}|^2.\!\!
\end{align}
In the last expression, I use that for $d \ll N$, the largest eigenvalue of $V$, i.e. $v_1$, will completely dominate. Now, we simply need to get the overlaps $|\braket{v_1|c_j}|^2$. It turns out that only $\braket{v_1|c_3}$ and $\braket{v_1|c_4}$ are nonzero. These yield
\begin{align}
C_z^{(1)} = |\braket{v_1|c_3}|^2 &= \frac{[v_1 - (e^{\beta|J|/4} + e^{-3\beta|J|/4})]^2}{[v_1 - (e^{\beta|J|/4} + e^{-3\beta|J|/4})]^2 + 4}, \nn \\
C_z^{(2)} = |\braket{v_1|c_4}|^2 &= \frac{4}{[v_1 - (e^{\beta|J|/4} + e^{-3\beta|J|/4})]^2 + 4}.
\end{align}
Since $Z_J = v_1^N$, we finally get
\begin{align}
C_{z}(d) =\, & v_1^{-d}\sum_{j} c_j^d |\braket{v_1|c_j}|^2 \nn \\
=\, & \frac{[v_1 - (e^{\beta|J|/4} + e^{-3\beta|J|/4})]^2}{[v_1 - (e^{\beta|J|/4} + e^{-3\beta|J|/4})]^2 + 4} \left[\frac{v_3}{v_1}\right]^d \nn \\
+ &\frac{4}{[v_1 - (e^{\beta|J|/4} + e^{-3\beta|J|/4})]^2 + 4} \left[\frac{v_4}{v_1}\right]^d \nn \\
=&\, C_{z}^{(1)} e^{-d/\xi_1(\beta J)} + C_{z}^{(2)} e^{-d/\xi_2(\beta J)}
\end{align}
giving a sum of two exponentially decaying terms. The correlation lengths are
\begin{align}
\!\!\!\xi_1(\beta J) = \left[\ln \left( \frac{v_1}{v_3} \right)\right]^{-1}\!\!, \; \xi_2(\beta J) = \left[\ln \left( \frac{v_1}{v_4} \right)\right]^{-1}\!\!.\!
\label{eq.correlation_lengths}
\end{align}
I note that $\xi_1 > \xi_2$ for any temperature. Inserting $v_1,v_3$ in the upper line leads to the expression in Eq. \eqref{eq.correlation_length} of the main text. 

\section{Disordered potential at infinite temperature} \label{app.disordered_potential}
Here, I derive the probability distribution of the magnetic potential $V(x)$. After $|x|$ hops, the possible values of the \emph{trans-leg} potential are
\begin{equation}
V_\perp(x) = n \frac{J}{2}, \; n\in \{-|x|, -|x| + 1, \dots, |x|\}.
\end{equation}
I want to calculate what the probabilities $P(V_\perp(x) = n J/2)$ are. To do so, the change in the potential may be described by the transition operator
\begin{equation}
\!\! T = \sum_{n} \left[\frac{1}{2}\ket{n}\bra{n} + \frac{1}{4}\ket{n+1}\bra{n} + \frac{1}{4}\ket{n-1}\bra{n} \right].\!\!
\end{equation}
Here, $\ket{n}$ denotes the outcome $n J/2$. The probability of $n J/2$ after $|x|$ hops is, therefore,
\begin{equation}
P\left(V_\perp(x) = n \frac{J}{2}\right) = \bra{n}T^{|x|}_\perp\ket{0}.
\end{equation}
To calculate this transition element, it is beneficial to use the eigenvectors of $T$. In particular, we let
\begin{equation}
\ket{n} = \frac{1}{\sqrt{N}}\sum_k e^{ikn}\ket{k}.
\end{equation}
Here, $k\in(-\pi, \pi]$. The transition operator is diagonal in these vectors
\begin{equation}
T_\perp = \sum_k t_k \ket{k}\bra{k},
\end{equation}
with $t_k = [1 + \cos(k)]/2 = \cos^2(k/2)$. Now, 
\begin{align}
&P\left(V_\perp(x) = n \frac{J}{2}\right) = \bra{n}T^{|x|}_\perp\ket{0} \nn \\
&= \sum_{k,q}\braket{n|q}\bra{q} T^{|x|}\ket{k}\braket{k|0} \nn \\
&= \sum_{k}\braket{n|k}t_k^{|x|}\braket{k|0} = \frac{1}{N}\sum_k e^{-ikn} t_k^{|x|}. 
\end{align}
We may turn this into an integral, yielding
\begin{align}
&P\left(V_\perp(x) = n \frac{J}{2}\right) = \frac{1}{N}\sum_k e^{-ikn} t_k^{|x|} \nn \\
&\to \int_{-\pi}^{\pi} \frac{dk}{2\pi} e^{-ikn} t_k^{|x|} = \int_{0}^{\pi} \frac{dk}{\pi} \cos(kn) t_k^{|x|}.
\end{align}
For any value of $n$ and $x$ this allows us to get the probabilities. Furthermore, we also compute the variance of the potential. Explicitly, 
\begin{align}
{\rm Var}[V_\perp(x)] &= \sum_{-|x|}^{|x|} P\left(V_\perp(x) = n \frac{J}{2}\right) \left(n \frac{J}{2}\right)^2 \nn \\
&= \frac{J^2}{4} \int_{0}^{\pi} \frac{dk}{\pi} \left[\sum_{-|x|}^{|x|} n^2 \cos(kn)\right] t_k^{|x|}.
\end{align}
The sum may be evaluated using Wolfram Alpha to yield
\begin{align}
&\sum_{-|x|}^{|x|} n^2 \cos(kn) = |x| \cos(k|x|) \left[\cot^2(k/2) + x + 1\right] \nn \\
& - \frac{1}{2}\cot(k/2)\sin(k|x|)\left[\cot^2(k/2) -2x^2 + 1\right].
\end{align}
Inserting this above gives the very simple result
\begin{align}
{\rm Var}[V_\perp(x)] = \frac{J^2}{8}|x|.
\end{align}
Since the frustation potential performs a (classical) random walk, the variance scales linearly in $|x|$.\\

Let us, equivalently, determine the probability distribution for the intra-leg potential $V_\parallel(x)$. I note that the change in this potential from site to site is equivalent to the transfer matrix
\begin{align}
T_\parallel = \frac{1}{4}\left[\ket{+1}\bra{0} + \ket{-1}\bra{0}\right] + \frac{1}{2}\sum_{n = -1}^{+1}\ket{n}\bra{n}.
\end{align}
To determine the probability distribution in this case, given by $\bra{n}T_\parallel^{|x|}\ket{0}$, we note that
\begin{align}
T_\parallel^{|x|}\ket{0} = \frac{1}{2}\ket{0} + \frac{1}{4}\left[\ket{+1} + \ket{-1}\right],
\end{align}
for any integer $x \neq 0$. So $P(V_\parallel(x) = 0) = 1/2$, and $P(V_\parallel(x) = \pm J / 2) = 1/4$ for any $x \neq 0$, whereby the variance is
\begin{align}
\!\!{\rm Var}[V_\parallel(x)] &= \!\!\sum_{n = -1}^{+1}\!\! P\left(V_\parallel(x) = n \frac{J}{2}\right) \left(n \frac{J}{2}\right)^2 = \frac{J^2}{8}.\!
\end{align}
Since the trans- and intraleg potentials are uncorrelated, their variances add
\begin{equation}
{\rm Var}[V(x)] = \frac{J^2}{8}\left[|x| + 1\right].
\end{equation}
This shows very explicitly that $V_\perp(x)$ dominates the distribution for large $|x|$. 

\section{Exact recursive solution} \label{sec.recursive_solution}
In the main text, I set up an effective Hamiltonian for a given spin realization and use an exact diagonalization (ED) package in Python to compute the dynamics from there. Here, I show that by going to the frequency domain, the equations of motion may be solved exactly. The required Fourier transformation to get the associated dynamics is, however, numerically heavier than actually using the ED package in Python.

By expressing the non-equilibrium wave function in terms of the retarded and advanced states \cite{Nielsen2022_1,Nielsen2022_2} $\ket{\Psi_\bsigma(\tau)} = \ket{\Psi^{\rm R}_\bsigma (\tau)} + \ket{\Psi^{\rm A}_\bsigma(\tau)} = \te^{-\eta|\tau|}[\theta(\tau)\ket{\Psi_\bsigma(\tau)} + \theta(-\tau)\ket{\Psi_\bsigma(\tau)}]$, I express the Schr{\"o}dinger equation, $i\partial_\tau \ket{\Psi_\bsigma(\tau)} = \hat{H}\ket{\Psi_\bsigma(\tau)}$ in frequency space
\begin{align}
(\omega + i\eta) \ket{\Psi^{\rm R}_\bsigma(\omega)} &= +i\ket{\Psi_\bsigma(\tau = 0)} + \hat{H} \ket{\Psi^{\rm R}_\bsigma(\omega)}.
\label{eq.schrodinger_equation_frequency_space}
\end{align}
Here, $\eta$ is a positive infinitesimal. Denoting the probability amplitudes of $\ket{\Psi^{\rm R}_\bsigma(\omega)}$ as $R_\bsigma(x,\omega)$ and using that the advanced state simply has the complex conjugated terms of the retarded state, $\ket{\Psi^{\rm A}_\bsigma(\omega)} = [\ket{\Psi^{\rm R}_\bsigma(\omega)}]^*$, then shows that $C_\bsigma(x,\tau)$ can be retrieved as the Fourier transform
\begin{equation}
C_\bsigma(x,\tau) = \int \frac{d\omega}{2\pi} \te^{-i(\omega + i\eta)\tau} \times 2\Re[R_\bsigma(x,\omega)].
\label{eq.C_x_and_R_x}
\end{equation}
Crucially, the amplitudes $R_\bsigma(x,\omega)$ satisfy a set of equations of motion, 
\begin{align}
[\omega + i\eta] R_\bsigma(x,\omega) =&\, i\delta_{x,0} + V_\bsigma(x) R_\bsigma(x,\omega) \nn \\
&+ t\left[R_\bsigma(x-1,\omega) + R_\bsigma(x+1,\omega)\right],
\label{eq.equations_of_motion_freq}
\end{align}
which may be solved recursively, as has been detailed recently in similar contexts \cite{Nielsen2022_1,Nielsen2023_1,Nielsen2023_2}. Here, $V_\bsigma(x)$ designates the magnetic potential experiences by the hole as it moves through the lattice. By finally defining the recursion function $f_\bsigma(x,\omega)$ through the relations
\begin{align}
R_\bsigma(x+1,\omega) &= t f_\bsigma(x+1,\omega) \, R_\bsigma(x,\omega), \; x \geq 0, \nn \\
R_\bsigma(x-1,\omega) &= t f_\bsigma(x-1,\omega) \, R_\bsigma(x,\omega), \; x \leq 0.
\label{eq.recursive_amplitudes}
\end{align}
leads to the recursive solutions 
\begin{align}
f_\bsigma(x,\omega) &= \frac{1}{\omega \!+\! i\eta \!-\! V_\bsigma(x) \!-\! t^2 f_\bsigma(x \!+\! 1,\omega)}, \; x > 0, \nn\\
f_\bsigma(x,\omega) &= \frac{1}{\omega \!+\! i\eta \!-\! V_\bsigma(x) \!-\! t^2 f_\bsigma(x \!-\! 1,\omega)}, \; x < 0.
\end{align}
Inserting this into the equations of motion for $x = 0$ yields the lowest order amplitude
\begin{align}
R_\bsigma(0,\omega) &= \frac{i}{\omega \!+\! i\eta \!-\! V_\bsigma(0) \!-\! t^2[f(-1,\omega) \!+\! f(1,\omega)]},
\end{align}
which may be identified simply as the retarded hole Green's function for the spin realization $\bsigma$. The higher-order amplitudes
\begin{align}
R_\bsigma(x,\omega) &= t^{x\phantom{||}}\!\!\prod_{j = +1}^x f_\bsigma(x,\omega) \times R_\bsigma(0,\omega), \; x > 0, \nn \\
R_\bsigma(x,\omega) &= t^{|x|}\!\!\prod_{j = -1}^x f_\bsigma(x,\omega) \times R_\bsigma(0,\omega), \; x < 0.
\label{eq.amplitudes_final}
\end{align}
are found by using the recursive structure in Eq. \eqref{eq.recursive_amplitudes}. Finally, using the Fourier transform in Eq. \eqref{eq.C_x_and_R_x}, $C_\bsigma(x,\tau)$ is found. 

\section{Appropriate sampling intervals in the Metropolis-Hastings algorithm} \label{sec.monte_carlo_sampling}
\begin{figure}[th!]
\begin{center}
\includegraphics[width=1.0\columnwidth]{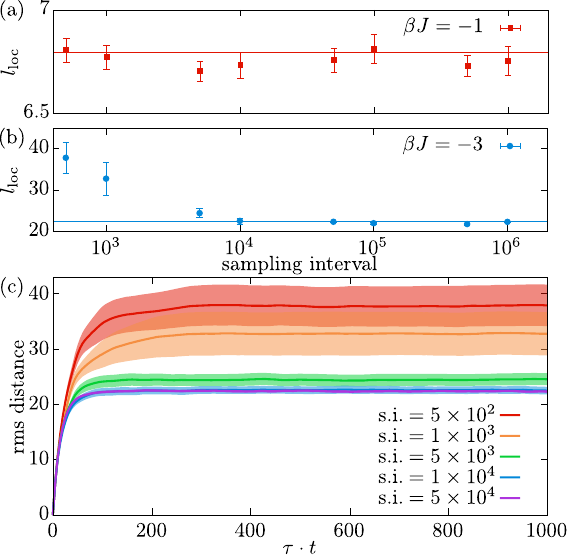}
\end{center}\vspace{-0.5cm}
\caption{Obtained localization length for $|J|/t = 2.5$ as a function of the sampling interval for (a) $\beta J = -1$ and (b) $\beta J = -3$ and compared to the value used in the main text (lines). The error bars show the estimated standard errors. In the high-temperature case (a), there is no perceived sensitivity to the sampling interval. In the low-temperature regime (b), however, an overly rapid sampling leads to an overestimation of the localization length. For sampling intervals above $10^4$ all points lie within $1\sigma$ from the value used in the main text (blue line). (c) The underlying rms dynamics for $\beta J = -3$, for varying sampling intervals (s.i.) is shown as a function of time $\tau$ in units of the hopping amplitude $t$. The shaded areas indicate the estimated standard error.}
\label{fig.sampling_accuracy} 
\vspace{-0.25cm}
\end{figure}

In this Appendix, I briefly investigate the sensititivity of the hole dynamics on how the sampling in the applied Metropolis-Hasting Monte Carlo algorithm is performed. To assess this, I compute the hole dynamics and the associated localization length for varying sampling intervals, i.e. the number of generated samples for every kept sample. This analysis is shown in Fig. \ref{fig.sampling_accuracy}. At high temperatures in Fig. \ref{fig.sampling_accuracy}(a), no perceived sensitivity to the sampling interval is seen. This is presumably because the autocorrelation time is much shorter than the investigated intervals. At low temperatues, Fig. \ref{fig.sampling_accuracy}(b), however, it is clearly seen that at too low intervals, the localization length is greatly overestimated. Finally, in Fig. \ref{fig.sampling_accuracy}(c) I plot the underlying rms dynamics for the low-temperature case for varying sampling intervals. This explicitly shows the dramatic decrease in the estimated statistical errors, as well as a convergence to a single well-defined line.

\bibliography{ref_thermal_localization}

\end{document}